\documentclass[journal=nalefd,manuscript=letter,layout=twocolumn]{achemso}

\usepackage[version=3]{mhchem} 
\usepackage{graphicx}
\usepackage{amsmath}
\usepackage{float}
\usepackage{amssymb}
\usepackage{color}
\usepackage[dvipsnames]{xcolor}
\usepackage{changes}
\usepackage{mathrsfs}
\usepackage{comment}
\usepackage{titlesec}
\titleformat{\subsection}[runin]{}{}{}{}[]

\title{Giant Spin Lifetime Anisotropy and Spin-Valley Locking in Silicene and Germanene from First-Principles Density-Matrix Dynamics}
\date{\today}

\author{Junqing Xu} \affiliation{Department of Chemistry and Biochemistry, University of California, Santa Cruz, CA 95064, USA}
\author{Hiroyuki Takenaka} \affiliation{Department of Chemistry and Biochemistry, University of California, Santa Cruz, CA 95064, USA}
\author{Adela Habib} \affiliation{Department of Physics, Applied Physics and Astronomy, Rensselaer Polytechnic Institute, 110 8th Street, Troy, New York 12180, USA}  
\author{Ravishankar Sundararaman}\email{sundar@rpi.edu} \affiliation{Department of Materials Science and Engineering, Rensselaer Polytechnic Institute, 110 8th Street, Troy, New York 12180, USA}
\author{Yuan Ping}\email{yuanping@ucsc.edu} \affiliation{Department of Chemistry and Biochemistry, University of California, Santa Cruz, CA 95064, USA}

\let\oldmaketitle\maketitle
\let\maketitle\relax
\begin{document}
\twocolumn[
\begin{@twocolumnfalse}
\oldmaketitle
\vspace{-0.1in}

\begin{abstract} Through First-Principles real-time Density-Matrix (FPDM) dynamics simulations, we investigate spin relaxation due to electron-phonon and electron-impurity scatterings with spin-orbit coupling in two-dimensional Dirac materials - silicene and germanene, at finite temperatures and under external fields. We discussed the applicability of conventional descriptions of spin relaxation mechanisms by Elliott-Yafet (EY) and D'yakonov-Perel' (DP) compared to our FPDM method, which is determined by a complex interplay of intrinsic spin-orbit coupling, external fields, and electron-phonon coupling strength, beyond crystal symmetry. For example, the electric field dependence of spin relaxation time is close to DP mechanism for silicene at room temperature, but rather similar to EY mechanism for germanene. Due to its stronger spin-orbit coupling strength and buckled structure in sharp contrast to graphene, germanene has a giant spin lifetime anisotropy and spin valley locking effect under nonzero $E_{z}$ and relatively low temperature. More importantly, germanene has extremely long spin lifetime ($\sim$100 ns at 50 K) and ultrahigh carrier mobility, which makes it advantageous for spin-valleytronic applications. \end{abstract} \end{@twocolumnfalse} \vspace{0.2in} ] \maketitle

\section*{Introduction}

Shortly after the discovery of graphene, significant advances have
been made in the field of spintronics, exploiting spin transport instead
of charge transport, with much less dissipation and unprecedented
potentials for low-power electronics. Several properties are key parameters
for optimizing spin transport, such as long spin lifetime and diffusion
length, high electronic mobility, spin-valley locking effect. Long
spin lifetime and diffusion length ensure robust spin state during
propagation in a device. The spin-valley locking effect is valuable
for the emerging field of research - "valleytronics"\cite{schaibley2016valleytronics,zhang2014generation,tao2020valley},
which utilizes the valley-pseudospin degree of freedom as basic unit
for quantum information technology.

Graphene is a very promising spintronic material\citep{Han_2014},
in particular, with the extremely high electronic mobility\citep{SarmaRMP11}
and the longest known spin diffusion length at room temperature\citep{drogeler2016spin}.
However, due to its weak spin-orbit coupling (SOC), spin-valley locking
effect may be realized only through external effects, e.g., through
proximity effect - by interfacing with large SOC materials such as
transition metal dichalcogenides (TMDs)\citep{cummings2017giantspinlifetime,GmitraPRL17}.
Other 2D materials\citep{avsar2020colloquium} including TMDs\citep{dey2017gate}
have also shown exciting properties for spin-/valley-tronics\cite{schaibley2016valleytronics,zhang2014generation,tao2020valley}.
For instance, in Ref. \citenum{dey2017gate}, the ultralong spin/valley
lifetime is observed (e.g. 2 $\mu$s in p-type monolayer WSe$_{2}$
at 5 K) and the spin lifetime is insensitive to in-plane magnetic
fields, which is a signature of the spin-valley locking effects. However,
in general, TMDs have much lower electric mobility than graphene\citep{cheng2018limits,ciccarino2018dynamics},
which is not ideal for electronic device applications.

As the counterparts of graphene, silicene and germanene have attracted
significant attention throughout a decade due to their resemblance
and distinction from graphene\citep{ezawa2012valley,tsai2013gated,D_vila_2014,zhang2016structural,Zhao_2016}.
They possess several remarkable properties, including gate-tunable
carrier concentration, high electric mobility\citep{Shao_2013,Ye_2014}
(Fig. S8), quantum spin hall effects\citep{CCliu11,Acun_2015}, etc.
Moreover, due to the buckled geometry, their SOC strength is highly
enhanced and they will also have the advantages of tunable band structures
by applying perpendicular electric fields $E_{z}$ and high spin polarizations,
which are promising for spintronic applications\citep{tsai2013gated}.
Their electronic structure under finite $E_{z}$ has similarity to
those of TMDs: the signs of spins in $K$ and $K'$ valleys being
opposite which is imposed by time-reversal symmetry, band splittings
induced by SOC and highly polarized states. Therefore, silicene and
germanene can combine the advantages of both graphene and TMDs but
avoid some shortcomings such as limited spin lifetimes ($\leq$ tens
of ns) in graphene samples\cite{han2011spin,kamalakar2015long}, the
absence of spin-valley locking in graphene, and low mobilities in
TMDs\citep{cheng2018limits,ciccarino2018dynamics}.

\begin{figure*}[!ht]
\includegraphics[scale=0.12]{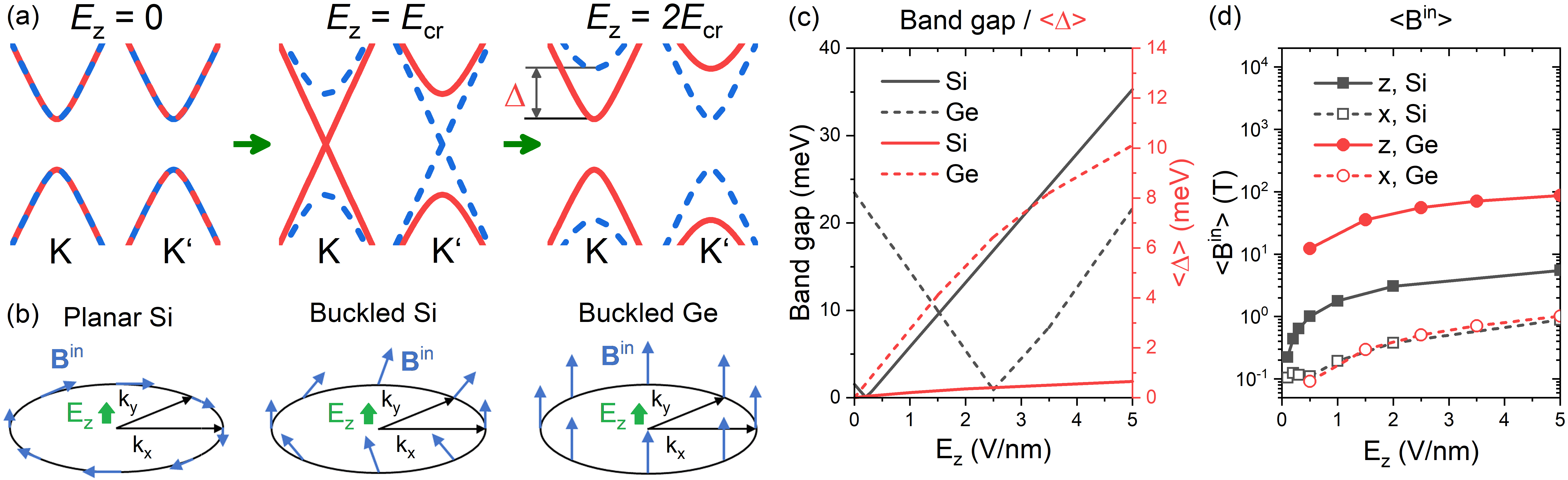}

\caption{Electronic quantities of silicene and germanene under different perpendicular
electric field ($E_{z}$). (a) The schematic diagrams of band structures
of silicene and germanene under different $E_{z}$ (see calculated
band structures in Fig.~S3-S5), where $E_{cr}$ is the critical electric
field leading to zero band gap. (b) The schematic diagrams of internal
magnetic fields ${\bf B}^{\mathrm{in}}$ (blue arrows) at a Fermi
circle near one Dirac cone of planar silicene, (buckled) silicene,
(buckled) germanene induced by breaking inversion symmetry under finite
$E_{z}$. ${\bf B}_{kn}^{\mathrm{in}}=2\Delta_{kn}{\bf S}_{kn}^{\mathrm{exp}}/\left(g_{e}\mu_{B}\right)$,
where $k$ and $n$ are k-point and band indices, respectively. $g_{e}\mu_{B}$
is the electron spin gyromagnetic ratio. $\Delta_{kn}$ is energy
difference between two conduction (valence) bands when $n$ is a conduction
(valence) band. ${\bf S}^{\mathrm{exp}}$ is a vector and ${\bf S}^{\mathrm{exp}}\equiv\left(S_{x}^{\mathrm{exp}},S_{y}^{\mathrm{exp}},S_{z}^{\mathrm{exp}}\right)$,
where $S_{i}^{\mathrm{exp}}$ is spin expectation value along direction
$i$ and is the diagonal element of spin matrix $s_{i}$ in Bloch
basis. (c) Band gaps (black lines) and the averaged band splitting
energies $\left\langle \Delta\right\rangle $ (red lines) between
two conduction/valence bands. (d) The averaged out-of-plane ($B_{z}^{\mathrm{in}}$)
and in-plane internal magnetic fields ($B_{x}^{\mathrm{in}}$). Throughout
this work, $\left\langle A\right\rangle $ means the average\cite{restrepo2012full}
of electronic quantity $A$ and $\left\langle A\right\rangle =\sum_{kn}f'\left(\varepsilon_{kn}\right)A_{kn}/\sum_{kn}f'\left(\varepsilon_{kn}\right)$.
$f'$ is the derivative of Fermi-Dirac distribution function. In this
figure, for averaging, $T=300$ K and chemical potential $\mu$ is
set in the middle of the band gap.\label{fig:electronic}}
\end{figure*}

Despite of their promising properties, the potential for spin-based
information technologies by silicene and germanene has yet to be demonstrated.
Understanding spin dynamics and transport in materials is of key importance
for spintronics and spin-based quantum information science, and one
key metric of useful spin dynamics is spin lifetime $\tau_{s}$. Compared
with graphene, of which $\tau_{s}$ has been extensively studied experimentally
and theoretically\citep{drogeler2016spin,cummings2016effects,habib2020electric},
$\tau_{s}$ of silicene and germanene have not been measured and the
existing few theoretical studies were done based on relatively simple
models\citep{bishnoi2013spin,babaee2019spin}, which do not include
realistic interactions with phonons and impurities. Recently we developed
a First-Principles Density-Matrix (FPDM) approach with quantum descriptions
of scattering processes between electron-phonon (e-ph), electron-impurities
(e-i) and electron-electron, to simulate spin-orbit mediated spin
dynamics in general solid-state systems with arbitrary symmetry\citep{xu2020ab,xu2020spin}.
By employing this new method, we will be able to predict $\tau_{s}$
of silicene and germanene at finite temperatures with realistic interactions
with environment without introducing any simplified model or empirical
parameters, for ps to $\mu$s timescale simulation.

\section*{Results and discussions}

\subsection*{\bf{Electronic structure.}}

We first show electronic quantities of silicene and germanene under
different $E_{z}$, which are closely related to spin dynamics and
essential for understanding spin relaxation mechanisms.

Fig.~\ref{fig:electronic}(a) describes a schematic picture of band
structures under $E_{z}$. DFT results of band gaps and the average
of band splittings $\left\langle \Delta\right\rangle $ between two
conduction/valence bands (broken Kramers' degeneracy under $E_{z}$
field) are shown in Fig.~\ref{fig:electronic}(c). '$\left\langle \right\rangle $'
represents taking average of electronic quantity $A$ by\cite{restrepo2012full}
$\left\langle A\right\rangle =\sum_{kn}f'\left(\epsilon_{kn}\right)A_{kn}/\sum_{kn}f'\left(\epsilon_{kn}\right)$,
where $f'$ is the derivative of Fermi-Dirac distribution function,
and $k$ and $n$ are k-point and band indices respectively. At $E_{z}=0$,
due to time-reversal and inversion symmetries, every two bands form
a Kramers degenerate pair.\citep{vzutic2004spintronics} A finite
$E_{z}$ splits a Kramers pair into spin-up and spin-down bands due
to broken inversion symmetry and the splitting increases with $E_{z}$
(see Fig.\ref{fig:electronic}(c)). As the earlier model Hamiltonian
study \citep{ezawa2012valley} presented, the phase transition from
topological insulators to band insulators happens at the critical
electric field $E_{cr}$ with the schematic band structures shown
in Fig.~\ref{fig:electronic}(a). Based on our DFT calculations,
the band gaps close in silicene and germanene (black solid and dashed
lines in Fig. \ref{fig:electronic}(c)) at 0.2 and 2.5 V/nm respectively,
and the gaps open again above $E_{cr}$.

The band splittings between spin-up and down under a finite $E_{z}$
are effectively induced by $k$- and band-dependent ``internal''
magnetic fields ${\bf B}^{\mathrm{in}}$, which are SOC fields induced
by broken inversion symmetry.\citep{vzutic2004spintronics} ${\bf B}^{\mathrm{in}}$
is defined as $2\Delta_{kn}{\bf S}_{kn}^{\mathrm{exp}}/\left(g_{e}\mu_{B}\right)$,
$g_{e}\mu_{B}$ is the electron spin gyromagnetic ratio. ${\bf S}^{\mathrm{exp}}\equiv\left(S_{x}^{\mathrm{exp}},S_{y}^{\mathrm{exp}},S_{z}^{\mathrm{exp}}\right)$,
where $S_{i}^{\mathrm{exp}}$ is spin expectation value along direction
$i$ and is the diagonal element of spin matrix $s_{i}$ in Bloch
basis.

In Fig.~\ref{fig:electronic}(b), we depict schematic diagrams of
${\bf B}^{\mathrm{in}}$ (blue arrows) on a Fermi circle near one
Dirac cone of planar silicene, buckled silicene, and buckled germanene
induced by finite $E_{z}$. In planar silicene, ${\bf B}^{\mathrm{in}}$
is purely in-plane, known as Rashba Spin-Orbit fields ${\bf B}_{\mathrm{R}}^{\mathrm{in}}$.
${\bf B}^{\mathrm{in}}$ in buckled silicene is analogous to an admixture
of ${\bf B}_{\mathrm{R}}^{\mathrm{in}}$ and out-of-plane field $B_{z}^{\mathrm{in}}$,
because buckled geometry results in the presence of $B_{z}^{\mathrm{in}}$.
Different from silicene, ${\bf B}^{\mathrm{in}}$ in buckled germanene
is nearly fully out-of-plane. This indicates that stronger intrinsic
SOC in germanene significantly increases out-of-plane $B_{z}^{\mathrm{in}}$
and, as a result, the proportion of in-plane component diminishes.
We examine the averaged $B^{\mathrm{in}}$ along z, $\left\langle B_{z}^{\mathrm{in}}\right\rangle $,
and along x, $\left\langle B_{x}^{\mathrm{in}}\right\rangle $, under
$E_{z}$ shown in Fig.\ref{fig:electronic}(d). While $\left\langle B_{x}^{\mathrm{in}}\right\rangle $
for silicene and germanene increase slowly with similar values, $\left\langle B_{z}^{\mathrm{in}}\right\rangle $,
for germanene due to stronger intrinsic SOC, rises much more rapidly
than that for silicene as a function of $E_{z}$. ${\bf B}^{\mathrm{in}}$
and its anisotropy between z and x are very important for spin relaxation.

\subsection*{\bf{Spin relaxation under finite $E_{z}$.}}

Spin lifetime, and even spin relaxation mechanism, can be tuned by
applying electric fields\cite{guimaraes2014controlling,habib2020electric}.
Understanding the effect of electric field is critical for the control
and manipulation of spin relaxation. 
we start our theoretical studies of spin relaxation from its electric
field dependence.

As we perform FPDM calculations of spin lifetimes in this work, it
is important to understand the connection and distinction between
FPDM method and previous theoretical models. Previously, spin relaxation
mechanisms are often analyzed based on phenomenological models, such
as Elliott--Yafet (EY) and D'yakonov-Perel' (DP) mechanisms\cite{vzutic2004spintronics}.
EY represents the spin relaxation pathway due to spin-flip scattering.
DP is activated when inversion symmetry is broken which results in
random spin precession between adjacent scattering events. We denote
their corresponding spin lifetime with $\tau_{s}^{\mathrm{EY}}$ and
$\tau_{s}^{\mathrm{DP}}$ respectively. They are often approximated
by some simplified relations\cite{fabian1998spin,leyland2007oscillatory,vzutic2004spintronics}:
(i) $\left(\tau_{s,i}^{\mathrm{EY}}\right)^{-1}\approx4\left\langle b_{i}^{2}\right\rangle \left\langle \tau_{p}^{-1}\right\rangle $
(EY relation) along direction $i$, where $\tau_{p}$ is carrier lifetime.
$b_{i}^{2}=1-2S_{i}^{\mathrm{exp}}$ is the degree of mixture of spin-up
and spin-down states, so called "spin mixing"\citep{vzutic2004spintronics,kurpas2019spin},
and is calculated at $E_{z}=0$. (ii) $\left(\tau_{s,i}^{\mathrm{DP}}\right)^{-1}\approx\left\langle \tau_{p}^{-1}\right\rangle ^{-1}\left\langle \text{\ensuremath{\Omega^{2}}}-\Omega_{i}^{2}\right\rangle $
(DP relation), where $\Omega_{i}=g_{e}\mu_{B}B_{i}^{\mathrm{in}}$
is Larmor precession frequency\citep{vzutic2004spintronics} with
$B_{i}^{\mathrm{in}}$ defined earlier. Another qualitative estimation
of the total spin relaxation rate by taking into account both mechanisms
is using\cite{bronold2002magnetic} $\left(\tau_{s}^{\mathrm{E+D}}\right)^{-1}=\left(\tau_{s}^{\mathrm{EY}}\right)^{-1}+\left(\tau_{s}^{\mathrm{DP}}\right)^{-1}$.
In the following, we will compare FPDM calculations and these phenomenological
models with first-principles input parameters including $\tau_{p}$,
$b_{i}^{2}$ and $\Omega_{i}$.

\begin{figure}[!ht]
\includegraphics[scale=0.35]{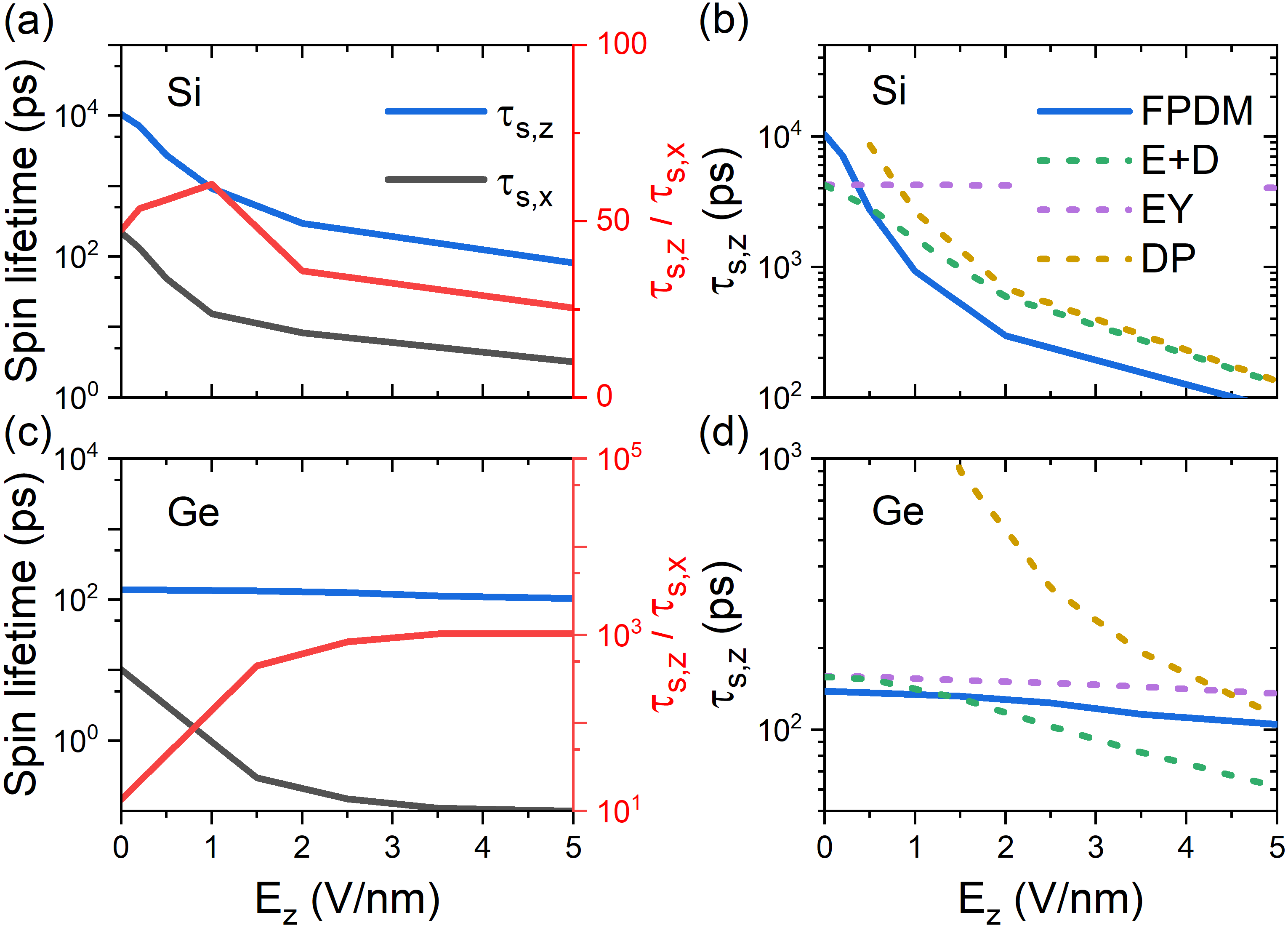}

\caption{Spin lifetimes $\tau_{s}$ (left $y$-axis) and its anisotropy (right
$y$-axis, red lines) of (a) intrinsic silicene and (c) intrinsic
germanene as a function of $E_{z}$ at 300 K. (b) and (d) are $\tau_{s,z}$
obtained by different methods of intrinsic silicene and germanene,
respectively. ``FPDM'' - first-principles real-time density-matrix
calculations. ``EY'' and ``DP'' correspond to $\tau_{s,z}^{\mathrm{EY}}$
and $\tau_{s,z}^{\mathrm{DP}}$ evaluated by EY and DP relations,
respectively. ``E+D'' corresponds to $\tau_{s}^{\mathrm{E+D}}=1/\left[1/\tau_{s}^{\mathrm{EY}}+1/\tau_{s}^{\mathrm{DP}}\right]$.\label{fig:300K_diff_Ez}}
\end{figure}

We first investigate out-of-plane and in-plane spin lifetime $\tau_{s,z}$
and $\tau_{s,x}$, respectively, and their anisotropy ($\tau_{s,z}/\tau_{s,x}$)
at $E_{z}=0$ and 300K. From Fig.\ref{fig:300K_diff_Ez} (a) and (b),
we find that (i) $\tau_{s,z}$ and $\tau_{s,x}$ of silicene are much
longer than germanene; (ii) large anisotropy (10-100) of $\tau_{s}$
is observed for both materials, e.g., 47 and 14 for silicene and germanene
at zero $E_{z}$ respectively, much greater than 0.5 for graphene\citep{habib2020electric,cummings2017giantspinlifetime}.
Both phenomena may be qualitatively understood based on EY relation
(typically dominant in inversion symmetric systems) as discussed in
the following. The comparison between FPDM calculations and EY relation
with first-principles inputs is shown in Fig.\ref{fig:300K_diff_Ez}
(c) and (d), and they give the same order of magnitude of spin lifetime.
Roughly speaking, the larger spin lifetime of silicene is mainly from
the smaller spin mixing $b^{2}$ compared with germanene based on
their intrinsic SOC strength. While the large spin anisotropy for
both systems is a result of large ratio of $b_{z}^{2}/b_{x}^{2}$
(Fig. S6(a)).

We then discuss $E_{z}$ dependence of spin relaxation. From Fig.~\ref{fig:300K_diff_Ez}(a)
and \ref{fig:300K_diff_Ez}(b), $\tau_{s,x}$ of silicene and germanene
rapidly reduces with increasing $E_{z}$. This trend can be qualitatively
understood as follows. Finite $E_{z}$ breaks inversion symmetry and
splits Kramers degeneracy. This induces ${\bf B}^{\mathrm{in}}$ with
a rapidly increased $z$ component as shown in Fig.~\ref{fig:electronic}(d),
and thus leads to fast in-plane spin relaxation (perpendicular to
$B_{z}^{\mathrm{in}}$) and reduces in-plane spin lifetime $\tau_{s,x}$.

Unlike $\tau_{s,x}$, out-of-plane spin lifetime $\tau_{s,z}$ of
germanene is insensitive to $E_{z}$ in Fig.~\ref{fig:300K_diff_Ez}(c)
although $\tau_{s,z}$ of silicene decreases fast with $E_{z}$(Fig.~\ref{fig:300K_diff_Ez}(a)).
To better understand $\tau_{s,z}$ dependence on $E_{z}$, we compare
FPDM $\tau_{s,z}$ with the model calculations based on EY ($\tau_{s,z}^{\mathrm{EY}}$),
DP ($\tau_{s,z}^{\mathrm{DP}}$) and EY+DP ($\tau_{s,z}^{\mathrm{E+D}}$)
mechanisms as introduced earlier, in Fig. \ref{fig:300K_diff_Ez}(b)
and \ref{fig:300K_diff_Ez}(d). From Fig.~\ref{fig:300K_diff_Ez}(b)
for silicene, we show that $\tau_{s,z}^{\mathrm{E+D}}$ and $\tau_{s,z}^{\mathrm{DP}}$
approximately agree with FPDM $\tau_{s,z}$ in trends. For germanene,
however, from Fig. \ref{fig:300K_diff_Ez}(d) we find that $\tau_{s,z}^{\mathrm{EY}}$
is in good agreement with the FPDM $\tau_{s,z}$, but neither $\tau_{s,z}^{\mathrm{E+D}}$
nor $\tau_{s,z}^{\mathrm{DP}}$ capture the qualitative trend. Therefore,
$z$-direction spin relaxation in germanene should be mostly driven
by EY mechanism, insensitive to $E_{z}$. The suppression of DP mechanism
in germanene under finite $E_{z}$ may be due to the huge ${\bf B}^{\mathrm{in}}$
anisotropy $B_{z}^{\mathrm{in}}/B_{x}^{\mathrm{in}}$ (see Fig. \ref{fig:electronic}(d)
and Fig. S6(b)): as $B_{z}^{\mathrm{in}}$ is so strong, any in-plane
spins will be quickly relaxed and all spins are pinned along $z$;
thus the total spin can only decay through direct spin-flip processes
but not through spin precession driven by ${\bf B}^{\mathrm{in}}$
which changes spin direction gradually.

\subsection*{\bf{Temperature dependence of $\tau_{s,z}$ and spin-valley locking.}}

\begin{figure}[!ht]
\includegraphics[scale=0.36]{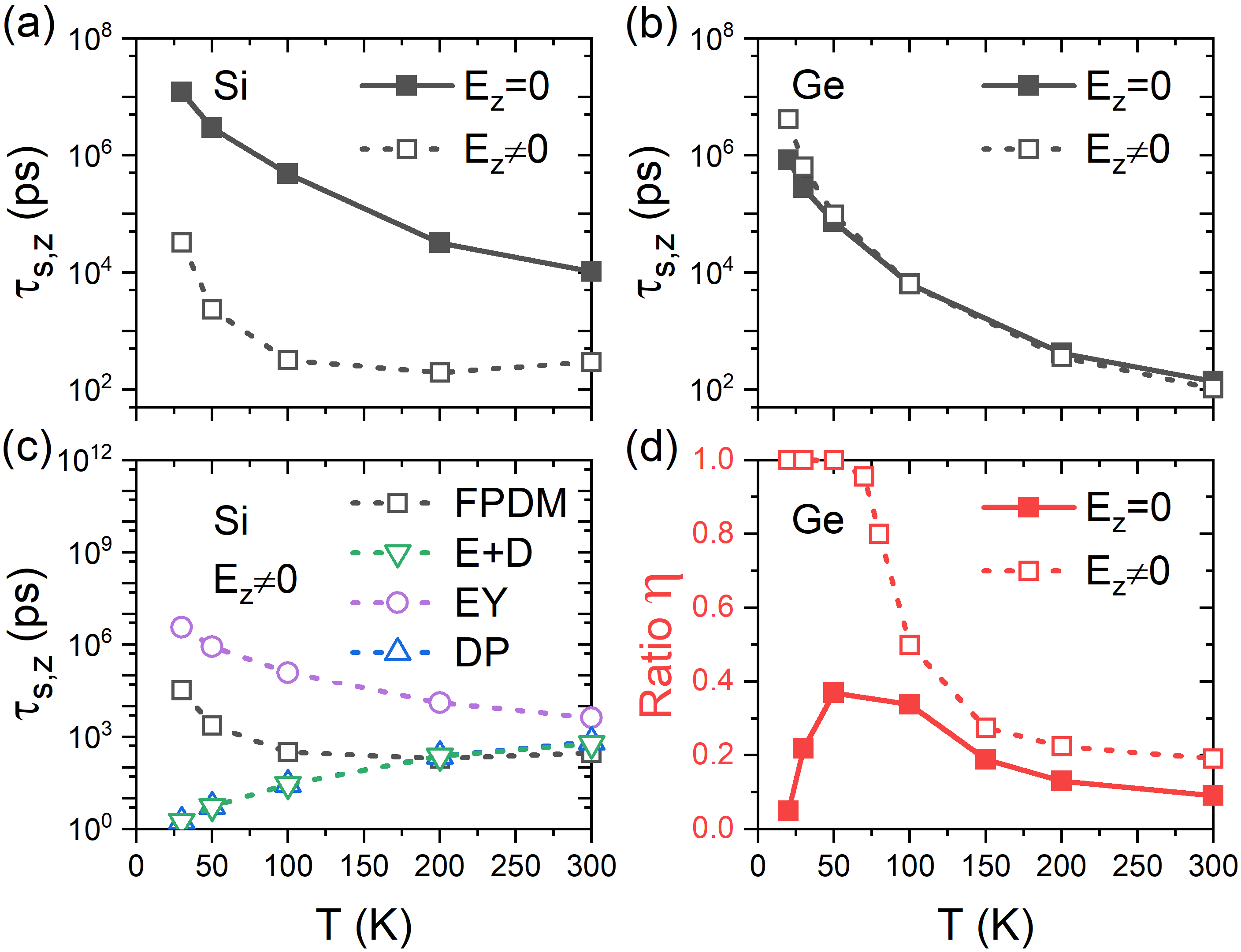}

\caption{Temperature dependent $\tau_{s,z}$ of intrinsic (a) silicene and
(b) germanene under $E_{z}=0$ and $E_{z}\protect\neq0$ (2 and 5
V/nm for silicene and germanene, respectively). (c) $\tau_{s,z}$
of intrinsic silicene under $E_{z}\protect\neq0$ obtained by different
methods. The labels of the curves have the same meanings as in Fig.~\ref{fig:300K_diff_Ez}.
(d) Relative intervalley spin relaxation contribution $\eta$ of germanene
under $E_{z}=0$ and $E_{z}\protect\neq0$. $\eta$ is defined as
$\eta=\frac{\left(\tau_{s,z}^{\mathrm{inter}}\right)^{-1}}{\left(\tau_{s,z}^{\mathrm{inter}}\right)^{-1}+w\left(\tau_{s,z}^{\mathrm{intra}}\right)^{-1}}$,
where $\tau_{s,z}^{\mathrm{inter}}$ and $\tau_{s,z}^{\mathrm{intra}}$
are intervalley and intravalley spin lifetimes, corresponding to scattering
processes between $K$ and $K'$ valleys and within a single $K$
or $K'$ valley, respectively. $\eta$ being close to 1 or 0 corresponds
to intervalley or intravalley scattering dominant spin relaxation,
respectively. See more details of $\eta$ in Ref. \citenum{fig3caption}.
\label{fig:temperature}}
\end{figure}

It is important to understand the sensitivity of spin lifetime to
temperature and determine the optimal operating temperature\cite{yang2015long,kikkawa1998resonant,xu2020spin}.
Therefore, we show temperature dependence of $\tau_{s,z}$ without
$E_{z}$ and with $E_{z}=$ 2 and 5 V/nm (higher than $E_{cr}$ by
$\sim$2 V/nm) in intrinsic silicene and germanene respectively in
Fig.~\ref{fig:temperature}(a) and (b). Without $E_{z}$, $\tau_{s,z}$
of both silicene and germanene increase fast on cooling. This is the
usual behavior of EY spin lifetime simply due to weaker e-ph
scattering with lowering temperature. Since phonon occupation is smaller
at a lower temperature, $\tau_{p}$ is longer (Fig.~S8), so $\tau_{s}$
is longer ($\tau_{s}\propto\tau_{p}$ with EY mechanism). Under finite
$E_{z}$, $\tau_{s,z}$ of germanene are similar to the values under
zero $E_{z}$ as shown in Fig. \ref{fig:300K_diff_Ez}(b), which are
expected from the discussions on $E_{z}$ dependence of germanene
above. In sharp contrast, finite $E_{z}$ significantly reduces $\tau_{s,z}$
of silicene and modifies the temperature dependence. To interpret
such complex temperature dependence, in Fig. \ref{fig:temperature}(c),
we compare FPDM $\tau_{s,z}$ and the model ones $\tau_{s,z}^{\mathrm{EY}}$,
$\tau_{s,z}^{\mathrm{DP}}$ and $\tau_{s,z}^{\mathrm{E+D}}$ for silicene.
All model relations fail to reproduce the temperature dependence under
finite $E_{z}$ for silicene. The failure of the DP relation in particular
below 200 K is probably because it is inapplicable for weak scattering
regime (weak e-ph scattering at low temperatures)\cite{vzutic2004spintronics}.
Other relations have been proposed for weak scattering\citep{vzutic2004spintronics,leyland2007oscillatory,wu2010spin},
e.g., $\tau_{s}\sim|\Omega|^{-1}$ and $\tau_{s}\sim2\tau_{p}$, but
none can capture the correct temperature dependence of $\tau_{s,z}$
of silicene under finite $E_{z}$, e.g., $\tau_{s}\sim|\Omega|^{-1}$
predicts $\tau_{s,z}<20$ ps at all temperatures investigated here.
Our theoretical studies highlight the importance of simulating spin
lifetime using FPDM method for reliable prediction of spin lifetimes
for large variations of external conditions.

\begin{figure*}[!ht]
\includegraphics[scale=0.32]{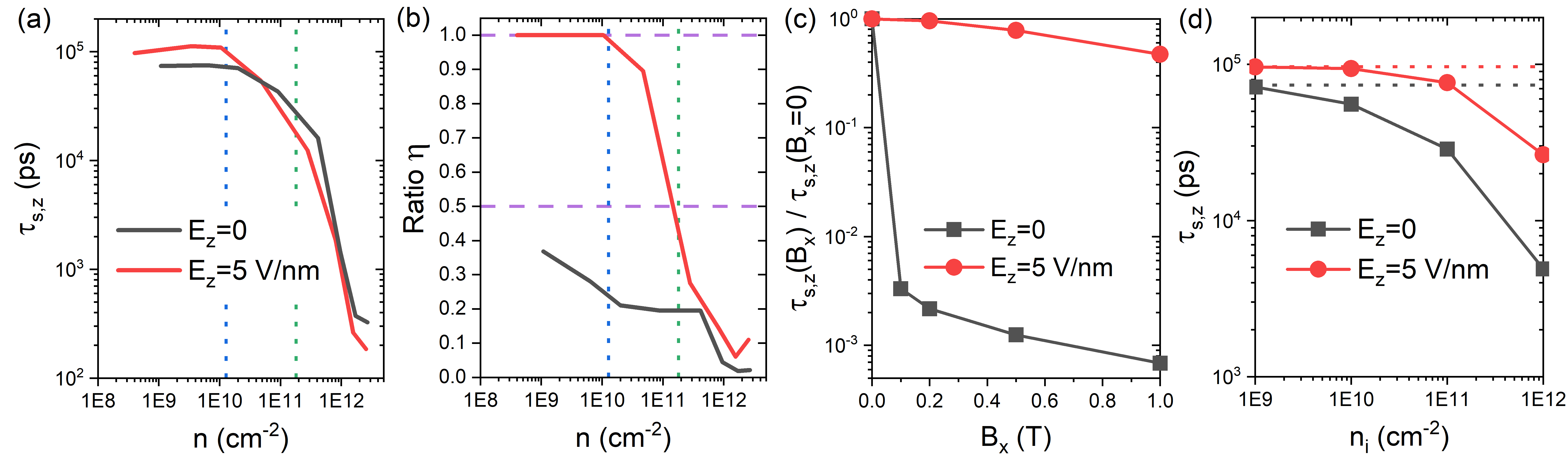}

\caption{Spin relaxation in germanene at 50 K under 0 and 5 V/nm. (a) $\tau_{s,z}$
and (b) relative intervalley spin relaxation contribution $\eta$
of germanene as a function of excess carrier density $n$, which is electron density $n_{e}$ minus hole density $n_{h}$, and controlled by chemical potential $\mu$. Positive carrier density
corresponds to electron doping. (c) $\tau_{s,z}$
of germanene as a function of in-plane magnetic field - $B_{x}$.
(d) $\tau_{s,z}$ of germanene as a function of impurity density $n_{i}$ of
neutral Ge vacancy (with $n_{e}$ and $n_{h}$ kept the same as intrinsic germanene). $\eta$ being close to 1 or 0 correspond
to spin relaxation being dominated by intervalley or intravalley scattering,
respectively. The blue and green dotted vertical lines correspond
to $\mu$ at the minima of first and second conduction bands respectively.
The black and red dotted lines are $\tau_{s,z}$ in the zero $n_{i}$ limit under $E_{z}=0$ and 5 V/nm,
respectively.\label{fig:50K_diff_n}}
\end{figure*}

Since the bands near the Fermi energy are composed of the Dirac cone
electrons around $K$ and $K'$ valleys in silicene and germanene,
spin relaxation process mostly arises from intervalley and intravalley
scatterings. To scrutinize contributions of intervalley and intravalley
scatterings to spin lifetime, we examine relative intervalley spin
relaxation contribution $\eta$ (see its definition in the caption
of Fig.~\ref{fig:temperature}). $\eta$ being close to 1 or 0 correspond
to spin relaxation being dominated by intervalley or intravalley scattering,
respectively. From Fig.~\ref{fig:temperature}(d), it is found that
for germanene under $E_{z}$=5 V/nm, the $\eta$ becomes close to
1 at $T\leq$70 K. This indicates that under these conditions spin
relaxation in germanene is dominated by intervalley process, which
is a signature of so-called spin-valley locking. In addition, the
long $\tau_{s,z}$ reaches around 100 ns at 50K and $E_{z}$=5 V/nm
and indicates stability of spin states against spin relaxation. Different
from germanene, in silicene, since the relatively small SOC splitting
even at $E_{z}$=2 V/nm as shown in Fig.~\ref{fig:electronic}(b),
the spin relaxation is mostly through intravalley scattering (not
shown), which indicates silicene will hardly show the spin-valley
locking property.

Formally the terminology ``spin-valley locking''
means that spin index (spin-up and spin-down) becomes locked with
the valley index (e.g., $K$ or $K'$)\cite{xu2014spin}. This implies
that (i) two valleys (e.g., $K$ and $K'$) exist with opposite spin
polarizations. (ii) Within one valley, spins are all highly polarized
along one direction, and carriers should have the same sign of spin.
This needs large SOC splitting, i.e., the spin-up and down bands being
largely separated.

Point (ii) will cause highly suppressed intravalley spin relaxation (i.e., relaxation through scattering processes within one valley).
We explain the reason using germanene
as an example: Under 5 V/nm, the SOC splitting for germanene is large,
23 meV at $K$. To have spins with the same sign, most
of carriers should be located around the band edges (which require low
temperatures and low carrier densities). Then the intravalley spin-flip
transition between an occupied state at band edges and an empty state at the second
conduction/valence band will be rather weak at relatively low temperatures,
since phonon occupation becomes negligible at the corresponding phonon
energy (comparable to SOC splitting).

On the other hand, for intervalley e-ph processes,
the corresponding phonon wavevectors are away from $\Gamma$ (e.g.,
around $K$) and the minimum phonon frequency is finite (e.g., 7 meV
for germanene, see Fig. S7). Therefore, with spin-valley locking,
spin relaxation through e-ph scattering (mostly intervalley) will
be highly suppressed leading to long spin lifetime, because of small
phonon occupations at relatively low temperatures.

\begin{table*}[!ht]
\begin{tabular}{|c|c|c|c|c|c|}
\hline 
T (K)  & $n_{i}$ (cm$^{-2}$)  & $\overline{\mu_{c}}$ (cm$^{2}$/V/s)  & $D$ (cm$^{2}$/s)  & $\tau_{s,z}$ (ns)  & $l_{||,s_{z}}$ ($\mu$m)\tabularnewline
\hline 
\hline 
300  & 0  & 3.2$\times$10$^{4}$  & 830  & 0.1  & 2.9\tabularnewline
\hline 
300  & 10$^{11}$  & 2.5$\times$10$^{4}$  & 620  & 0.1  & 2.5\tabularnewline
\hline 
50  & 0  & 3.8$\times$10$^{6}$  & 16700  & 97  & 400\tabularnewline
\hline 
50  & 10$^{11}$  & 4.5$\times$10$^{5}$  & 2000  & 76  & 120\tabularnewline
\hline 
50  & 10$^{12}$  & 5.8$\times$10$^{4}$  & 250  & 26  & 25\tabularnewline
\hline 
\end{tabular}

\caption{Spin dynamic and transport properties of intrinsic germanene under $E_{z}$=5 V/nm without and with neutral impurities (with impurity density $n_i$). $\overline{\mu_{c}}$
is the average of electron and hole mobility. The method of calculating
mobility is given in Supporting Information. The theoretical results
of electron and hole mobility are given in Fig. S8. $D$ is diffusion
coefficient. $l_{||,s_{z}}$ is spin diffusion length of $z$-direction
spin. The formula computing $D$ and $l_{||,s_{z}}$ are given in
the main text.\label{tab:spin_transport}}
\end{table*}

The existence of spin-valley locking not only leads to
long spin lifetime but also allows the utilization of previously developed
valleytronic technologies to design germanene-based devices\citep{schaibley2016valleytronics,tao2020valley}.
Our calculations provide guidance on the necessary conditions to realize spin-valley locking in germanene.

\subsection*{\bf{Carrier density and magnetic field dependence of $\tau_{s,z}$ for Germanene at low T.}}

As intervalley scattering was shown being dominant in germanene at
50 K with a very long spin lifetime, we further investigate its spin
relaxation as a function of carrier density, which
can be easily tuned by electrical gate experimentally. We will also
determine the range of carrier density where spin-valley locking happens.

As shown in Fig.~\ref{fig:50K_diff_n}(a), we plot $\tau_{s,z}$ of germanene as a function of excess carrier density $n$ (which is electron density $n_{e}$ minus hole density $n_{h}$, and controlled by chemical potential $\mu$).
We found it is very sensitive to $n$ at the degenerate doping range ($n\gtrsim$4$\times$10$^{10}$
cm$^{-2}$, corresponding to $\mu$ above the conduction
band minimum). As the carriers contributing to spin relaxation have
higher energies when $n$ increases, the strong $n$ dependence of
$\tau_{s,z}$ should be mainly a result of the strength of the e-ph
scattering being enhanced at higher energies (Fig. S10).

Moreover, we find that the relative intervalley spin relaxation contribution
$\eta>$0.9 at $n\leqslant4\times10^{10}$ cm$^{-2}$ under $E_{z}$=5
V/nm in Fig. \ref{fig:50K_diff_n}(b). This indicates spin relaxation
being dominated by intervalley processes and the presence of spin-valley
locking at the corresponding condition, consistent
with our above discussions about spin-valley locking.

We then investigate effects of spin-valley locking under in-plane
magnetic-field $B_{x}$. From Fig.~\ref{fig:50K_diff_n}(c) at 50K
for germanene, $B_{x}$ has weak effects on spin lifetime of germanene
under finite $E_{z}$. This is because the applied external magnetic
field is too weak compared with internal B field ${\bf B}^{\mathrm{in}}$
under finite $E_{z}$. This weak $B_{x}$ dependence of spin lifetime
is often used as an experimental evidence of spin-valley locking\cite{dey2017gate}.

\subsection*{\bf{Impurity effects and spin diffusion length.}}

Finally, we investigate the effects of the electron-impurity scattering.
As an initial theoretical investigation, we will consider only one
common neutral defect in germanene in this work - single Ge atom vacancy\cite{hastuti2019stability}.
From Fig.~\ref{fig:50K_diff_n}(d), we observe that $\tau_{s,z}$ are reduced by introducing impurities and the reduction becomes significant
when impurity density $n_{i}$ approaches 10$^{12}$ cm$^{-2}$. Another interesting
observation is that under finite $E_{z}=5$ V/nm, $\tau_{s,z}$ reduces
much less than the one with $E_{z}=0$. Our simulations suggest that
if $n_{i}$ can be controlled below 10$^{12}$ cm$^{-2}$,
especially under a finite $E_{z}$, germanene can exhibit long spin
lifetime over 100 ns at or below 50 K.

At the end, we compute in-plane spin diffusion length $l_{||,s_{z}}$
for $z$-direction (out-of-plane) spin polarization of germanene using
the relation\cite{vzutic2004spintronics} $l_{||,s_{z}}=\sqrt{D\tau_{s,z}}$,
where $D$ is diffusion coefficient. $D$ can be estimated using the
general form of Einstein relation\citep{kubo1966fluctuation}
$D=\mu_{c}(n_{e}+n_{h})/\frac{d(n_{e}+n_{h})}{d\mu}$, where $\overline{\mu_{c}}$
is the average of electron and hole mobility, which are obtained from solving Boltzmann
equation as detailed in the method section. From Table \ref{tab:spin_transport},
we can see $l_{||,s_{z}}$ of germanene at 300 K is 2-3 $\mu$m, shorter
than the longest measured value of graphene samples, $\sim$12 $\mu$m\cite{drogeler2016spin}.
At 50 K, as mobilities are higher and $\tau_{s,z}$ are longer, $l_{||,s_{z}}$
become 400 $\mu$m without impurities and 120 $\mu$m with $n_{i}=$10$^{11}$ cm$^{-2}$, which are much longer than experimental
values of graphene at different temperatures ranging from 1 to 40
$\mu$m\cite{avsar2020colloquium}.

\section*{Conclusions}

By employing our newly developed first-principles density-matrix dynamics
approach, we computed spin lifetime of two Dirac 2D materials - silicene
and germanene as a function of temperature, electrical doping and
neutral impurities, as well as applied electric and magnetic fields.
We find silicene and germanene have qualitative different spin relaxation
mechanisms under finite $E_{z}$ fields. 
We did systematic comparisons between our FPDM $\tau_{s}$ and those
estimated by phenomenological models with first-principles input parameters.
We find that germanene out-of-plane spin relaxation can be qualitatively
understood by EY relation, regardless of the one with and without
E field. On the other hand, spin relaxation in silicene is more complicated:
although at room temperature, the trends of $E_{z}$ dependence of
silicene $\tau_{s,z}$ can be captured by a combination of EY and
DP relation, the temperature dependence of $\tau_{s,z}$ under finite
$E_{z}$ cannot be explained by any simplified model relations.

We demonstrated giant spin lifetime anisotropy (two order of magnitude
higher than graphene) and provided the condition for spin-valley locking
with long spin lifetime in germanene. Specifically, we show that at
a low T - 50 K, $\tau_{s}$ of germanene can reach 100 ns and $l_{s}$
can exceed 100 $\mu$m (longer than graphene samples), if impurity
density is controlled low ($\leq$10$^{11}$ cm$^{-2}$).
This is very promising because the spin-valley locking property has
only been realized in either TMDs which usually have much lower carrier
mobility, or graphene on substrates which have complexity of interfacial
engineering. The realization of spin-valley locking in single materials
with long spin lifetime and ultrahigh mobility opens up highly promising
pathways for spin-valleytronic applications.

\section*{Methods}

To predict spin relaxation time from first principles, we employ our
newly developed \emph{ab initio} density-matrix dynamics approach,
which includes quantum descriptions of various scattering processes
and is applicable to general solid-state systems.\citep{xu2020ab,xu2020spin}
The density matrix master equation due to the e-ph and e-i scattering
in interaction picture reads: 
\begin{align}
\frac{d\rho_{12}\left(t\right)}{dt}= & \frac{1}{2}\sum_{345}\left\{ \begin{array}{c}
\left[I-\rho\left(t\right)\right]_{13}\rho_{45}\left(t\right)\times\\
\left[P_{32,45}^{\mathrm{e-ph}}\left(t\right)+P_{32,45}^{\mathrm{e-i}}\left(t\right)\right]\\
-\left[I-\rho\left(t\right)\right]_{45}\rho_{32}\left(t\right)\times\\
\left[P_{45,13}^{\mathrm{e-ph}}\left(t\right)+P_{45,13}^{\mathrm{e-i}}\left(t\right)\right]^{*}
\end{array}\right\} \nonumber \\
 & +H.C.,\label{eq:Lindblad}
\end{align}

where $\rho$ is density matrix. H.C. is Hermitian conjugate. The
subindex, e.g., ``1'' is the combined index of k-point and band.
The weights of k points must be considered when doing sum over k points.
$P^{\mathrm{e-ph}}$ and $P^{\mathrm{e-i}}$ are the generalized scattering-rate
matrices for the e-ph and e-i scattering respectively. Note that $P^{c}$
with $c$ being a scattering channel is related to its value $P^{S,c}$
in the Schrodinger picture as $P_{1234}^{c}\left(t\right)=P_{1234}^{S,c}\mathrm{exp}\left[i\left(\varepsilon_{1}-\varepsilon_{2}-\varepsilon_{3}+\varepsilon_{4}\right)t\right].$
$P^{S,c}$ is time independent and is computed from corresponding
e-ph or e-i matrix elements and electron and phonon energies.

All energies and matrix elements are calculated on coarse $k$ and
$q$ meshes using the DFT software JDFTx,\cite{sundararaman2017jdftx}
and are then interpolated to extremely fine meshes in a basis of maximally
localized Wannier functions.\cite{marzari1997maximally,PhononAssisted,NitrideCarriers}
Starting from at an initial state with a net spin, we evolve the density
matrix $\rho\left(t\right)$ through the master equation Eq. \ref{eq:Lindblad}
for a long enough simulation time, typically from ns to $\mu$s. We
then obtain the evolution of spin observable $S_{i}\left(t\right)$
($i=x,y,z$) from $\rho\left(t\right)$ (Eq. S1). At the end, spin
lifetime $\tau_{s,i}$ is obtained by fitting $S_{i}\left(t\right)$
to an exponential decay curve with decay constant $\tau_{s,i}$.

More details are given in Supporting Information Sec. SI and SII and
Ref. \citenum{xu2020ab}.

Using the same first-principles electron and phonon energies and matrix
elements on fine meshes, we calculate the carrier mobility by solving
the linearized Boltzmann equation using a full-band relaxation-time
approximation\citep{ciccarino2018dynamics} (Supporting Information
Sec. SVI).

\section*{Author contributions}

J.X. and H.T. performed the \emph{ab initio} calculations and analyses.
R.S. and Y.P. designed and supervised all aspects of the study. All
authors contribute to the writing of the manuscript.

\section*{Acknowledgements}

We thank Mani Chandra for helpful discussions. This work is supported by the Air Force Office of Scientific Research
under AFOSR Award No. FA9550-YR-1-XYZQ and National Science Foundation
under grant No. DMR-1956015. A. H. acknowledges support from the American
Association of University Women(AAUW) fellowship program. This research
used resources of the Center for Functional Nanomaterials, which is
a US DOE Office of Science Facility, and the Scientific Data and Computing
center, a component of the Computational Science Initiative, at Brookhaven
National Laboratory under Contract No. DE-SC0012704, the lux supercomputer
at UC Santa Cruz, funded by NSF MRI grant AST 1828315, the National
Energy Research Scientific Computing Center (NERSC) a U.S. Department
of Energy Office of Science User Facility operated under Contract
No. DE-AC02-05CH11231, the Extreme Science and Engineering Discovery
Environment (XSEDE) which is supported by National Science Foundation
Grant No. ACI-1548562 \citep{xsede}, and resources at the Center
for Computational Innovations at Rensselaer Polytechnic Institute.

\bibliography{main}

\end{document}